# Racism, Resistance, and Reddit: How Popular Culture Sparks Online Reckonings

Sherry Mason
Tawfiq Ammari

## Abstract

This study examines how Reddit users engaged with the racial narratives of *Lovecraft Country* and *Watchmen*, two television series that reimagine historical racial trauma. Drawing on narrative persuasion and multistep flow theory, we analyze 3,879 Reddit comments using topic modeling and critical discourse analysis. We identify three dynamic social roles—advocates, adversaries, and adaptives—and explore how users move between them in response to racial discourse. Findings reveal how Reddit's pseudonymous affordances shape role fluidity, opinion leadership, and moral engagement. While adversaries minimized or rejected racism as exaggerated, advocates shared standpoint experiences and historical resources to challenge these claims. Adaptive users shifted perspectives over time, demonstrating how online publics can foster critical racial learning. This research highlights how popular culture and participatory platforms intersect in shaping collective meaning-making around race and historical memory.

## Introduction

In the post-civil rights era, racism is often framed as a vestige of the past, promoting the illusion of a colorblind society and obscuring ongoing racial injustices (Bonilla-Silva, 2021). While Jim Crow racism justified overt segregation and racial hierarchy through biological essentialism, contemporary "colorblind racism" attributes inequality to nonracial cultural and economic factors. Today's systemic inequities are perpetuated through covert practices, policies that indirectly target marginalized communities, and daily interactions, as reflected in anti-CRT legislation, the rollback of DEI initiatives, ongoing police violence, and the resurgence of white nationalism (Onwuachi-Willig, 2020; Kim, 2021; Alfonseca, 2022; Elassar, 2025).

The suppression of education about historical atrocities such as the Tulsa Race Massacre (TRM), the Red Summer, and sundown towns (Kim, 2021; Alfonseca, 2022) further illustrates how American narratives of innocence and exceptionalism (Sturken, 2007; Delgado & Stefancic, 2023) are maintained through historical amnesia. Popular culture, particularly science fiction and horror, offers an important site for resurfacing these buried histories and facilitating public reckoning (Harper & Jenkins, 2022; Gillespie, 2020). Shows like *Lovecraft Country* and *Watchmen* reimagine suppressed racial traumas, challenging dominant racial ideologies (Grayson, Davies, & Philpott, 2009; Hall, 2006) and encouraging audiences to critically engage with the past. To negotiate and reconcile the meaning of such histories, audiences 'come together' on digital platforms like Reddit over shared interests—such as TV shows, where they can engage in discussions, express their opinions and beliefs, and form connections (Eriksson, 2016; Oddny, Ainslie, Lakshman, & Nathan, 2023). Reddit also serves as a contemporary site of racial discourse where masked racism resurfaces through moral disengagement and colorblind frames (Eschmann, 2023; Bonilla-Silva, 2021; Sue, 2006). Its pseudonymous structure enables both the amplification of racist ideologies and the creation of counterpublics through resource sharing, standpoint storytelling, and collective sensemaking (Ammari, Schoenebeck, & Romero, 2018; Massanari, 2015; Bruns & Burgess, 2015; Milner & Philips, 2020).

Building on narrative persuasion (Green & Brock, 2000; Cohen, Tal-Or, & Mazor-Tregerman, 2015) and multistep flow theory (Ognyanova, 2017; Kilgo et al., 2016), this study analyzes how Reddit users engaged with the racial narratives in *Lovecraft Country* and *Watchmen*, and how opinion leadership shapes discourse in pseudonymous online publics. We identify fluid social roles—advocates, adversaries, and adaptives—and examine how digital affordances influence processes of meaning-making around race and historical memory.

## Related Work

### Background

In 1921, a prosperous Black neighborhood known as "Black Wall Street," in the Greenwood district of Tulsa, Oklahoma, was violently attacked by a white mob, destroying homes, businesses, and taking an estimated 300 lives (Albright et al., 2021). This followed a series of terror and attacks on Black communities across the US at the turn of the twentieth century that would become known as the Red Summer (1919) "because it was so bloody" (McWhirter, 2011, p. 13). Such violence continued into the 1950s and 1960s with such events as the Trumbull Park Riots (1953-1966)-- a near decade long upheaval triggered by a Black family moving into a white residence in the Northern US (Hirsch, 1995) and sundown towns (1890 to 1968)—counties or municipalities in which segregation was enforced through curfew legislation and police action that permitted the harm or murder of Black people (as well as other people of color) who were in public after sunset (Lewis, 2021). The suppression of histories such as the TRM, Red Summer, Trumball Park, and sundown towns, among others, in formal education illustrates how state violence and racial trauma remain underacknowledged. The wave of legislation banning the teaching of Critical Race Theory across multiple states (Kim, 2021; Alfonseca, 2022) reflects a broader backlash against efforts to reckon with America's racial past, challenging national myths of innocence and exceptionalism (Sturken, 2007; Delgado & Stefancic, 2023). While Jim Crow racism justified racial hierarchy through overt segregationist practices, today's "colorblind racism" explains inequality as the outcome of nonracial dynamics, attributing disparities to cultural and economic factors (Bonilla-Silva, 2021).

Contemporary racial inequality persists through policies and laws that indirectly target marginalized groups, as well as through covert daily interactions both in-person and online (Bonilla-Silva, 2021; Eschmann, 2023). Prior research has examined how racial narratives circulate through digital spaces, focusing on both the dynamics of online discourse and the affordances of social media platforms (Eschmann, 2023; Massanari, 2015). The current study builds on this work by analyzing how Reddit users engage with racial narratives in *Lovecraft Country* and *Watchmen*, and how opinion leadership shapes these discussions. We identify two fluid roles among participants—advocates, who promote racial reckoning, and adversaries, who resist or undermine it—and examine how narrative persuasion and Reddit's platform features influence the flow of information and discourse around race.

### Race in Popular Culture

Popular culture, particularly science fiction and horror, offers audiences a way to engage with social inequities by exposing "the tension between Black lives as they are right now...and what they might be with a redistribution of power" (Harper & Jenkins, 2022, p. 78). Shows like *Watchmen* and *Lovecraft Country* serve as forms of public pedagogy, educating viewers on buried histories such as the Tulsa Race Massacre and the Red Summer. Popular culture not only reflects existing ideologies but also actively shapes them, operating as a battleground for cultural hegemony (Grayson, Davies, & Philpott, 2009; Hall, 2006). Historical examples like Harriet Beecher Stowe's *Uncle Tom's Cabin* (1852) and Harper Lee's *To Kill a Mockingbird* (1960) reveal how fictional narratives can both advance social causes and reinforce stereotypes (Jamieson, 2018; Moore & Pierce, 2007). Fictional media simplifies complex realities yet enables emotional engagement and fosters empathy (Green, Bilandzic, Fitzgerald, & Paravati, 2020; Mar & Oatley, 2008; Mar, Oatley, & Peterson, 2009). Today's television series increasingly bring suppressed histories into public view, offering critical counter-narratives to dominant racial ideologies.

*Watchmen* and *Lovecraft Country* exemplify how historical trauma is reimagined in popular culture. *Watchmen* recontextualizes Tulsa's 1921 racial violence within an alternate contemporary reality, while *Lovecraft Country* blends horror with historical references such as sundown towns and the Negro Motorist Green Book (Gillespie, 2020; Harper & Jenkins, 2022). Both shows challenge white racial framing (Slakoff, Douglas, & Smith, 2023) and prompted public discussions, particularly as audiences processed related events like the murder of George Floyd (Coffman & Schumar, 2023). Within today's convergence culture—where audiences remix and reinterpret media across platforms (Jenkins, 2006; Mikos, 2017)—fictional narratives around race contribute to collective meaning-making.

### *Digital Publics and Social Media Affordances*

Digital spaces like Reddit illustrate how racism adapts to new media environments. Drawing on Goffman's Dramaturgical Theory, Eschmann (2023) explains that overt racist rhetoric is now often masked in public discourse but can surface digitally. Reddit facilitates covert racial ideologies, contributing to the resurgence of "masked" racism. Modern racism increasingly manifests through microaggressions and colorblind ideologies that maintain systemic inequalities (Bonilla-Silva, 2021; Sue, 2006). Harmful acts are rationalized through moral disengagement processes (Bandura, 1999), and pseudonymity on Reddit can encourage this disinhibition (Ammari, Schoenebeck, & Romero, 2018).

Despite risks, social media also fosters positive engagement by offering safer spaces for marginalized users to share experiences and challenge dominant narratives (Corkum & Shead, 2023; Eschmann, 2023). Reddit, with its pseudonymous structure, sustains persistent user identities (Ammari, Schoenebeck, & Romero, 2018) and functions as a key space for political discourse, especially among younger, predominantly white, politically engaged users (Roozenbeek & Palau, 2017).

Audience behavior on Reddit reflects confirmation biases—users often seek content that reinforces pre-existing beliefs (Knobloch-Westerwick, Westerwick, & Johnson, 2015; Marchal, 2020)—yet online spaces still allow for interaction across political divides. Reddit's design has also been criticized for enabling toxic technocultures (Massanari, 2015). In digital environments, self-presentation becomes strategic: users manage their identity based on an "imagined audience" (Litt, 2012). Reddit's pseudonymity encourages both more "authentic" expression and disinhibited behaviors like trolling (Renninger, 2014; Marchal, 2020). The platform acts as a networked public (Thimm, 2017), where users form "issue publics" around topics like racial justice or environmental activism (Bruns & Burgess, 2015; Milner & Philips, 2020). Pseudonymity enables flexible audience management (Massanari, 2015) and supports both dominant discourse and the rise of counterpublics (Warner, 2002). Sense-making about racial histories on Reddit is thus shaped not just by user biases, but by the platform's technological affordances, participatory culture, and the dynamic complexity of digital publics.

*Multistep Flow Theory and Narrative Persuasion*

Narrative persuasion research suggests that cognitive and emotional absorption through elements like transportation (losing self-awareness and getting lost in a story) or character identification (when an audience takes on the perspective of a character) can make audiences less resistant to media messages (Green & Brock, 2000; Cohen, Tal-Or, & Mazor-Tregerman, 2015). However, audiences actively interpret content, accepting or rejecting elements based on personal experiences and ideologies, leading to the formation of interpretive communities (Boisvert & Gagnon, 2024).

Traditionally, two-step flow theory posited that media messages flow from outlets to opinion leaders, who then influence wider audiences. However, with the rise of digital platforms, multistep flow theory better captures how media messages are reshaped, amplified, and redistributed through multiple layers of interpersonal engagement (Ognyanova, 2017). Social media allows users not only to share but also to reinterpret and challenge narratives, creating new flows of influence at each stage. On Reddit, the structure of opinion leadership shifts. Rather than formal authority figures, opinion leaders emerge based on reputation metrics like upvotes and the persistence of recognizable pseudonyms (Kilgo et al., 2016). In these pseudonymous spaces, opinion leadership remains influential, suggesting that even anonymous or semi-anonymous users can significantly shape discourse and narrative meaning within digital publics.

**Methods**

We based our analysis on the methodological framework outline in Tornberg and Tornberg (2016) where computational methods (topic modeling) are used in tandem with critical discourse analysis to study large corpora. Specifically, topic models allow researchers to engage in a guided sampling strategy for critical discourse analysis.

*Data Collection*

We collected Reddit posts related to the HBO series *Lovecraft Country* (r/LovecraftCountry) and *Watchmen* (r/Watchmen) from December 1st 2019 to December, 31st 2022. An initial scrape yielded 7,601 threads. After topic filtering (described below), the final dataset included 1,764 threads.

*Topic Modeling*

We used BERTopic (Grootendorst, 2022) to identify major themes in Reddit discussions, treating each thread as a document. To explore model variation, we trained 25 topic models by varying the HDBSCAN minimum cluster size from 15 to 255 (increments of 10). Models were evaluated by coherence scores and topic granularity. Smaller models had higher coherence but fewer, less rich topics. We selected the model with a coherence score of 0.601 and minimum cluster size of 45, yielding 81 topics. Our analysis focused on threads prominently featuring at least one of five race-related topics. We focused our analysis on threads in which at least one of five race-related topics was prominently discussed. The topic names, associated keywords, and brief descriptions are presented below. The topic name, keywords, and a short description are presented below.

**Figure 1**

*Topics Related to Race for Subsequent Analysis*

| Topic Name | Topic Keywords | Short Description |
| --- | --- | --- |
| Rorschach, Racism, and Character Interpretation in Watchmen | rorschach, character, moore, racist, white, world, think, people | Critique and discussion of Rorschach's moral symbolism and racial implications in *Watchmen*. |
| Race and Policing in the U.S. | police, black, white, cops, racism, history, people, think | Conversations around racial injustice, policing, and systemic racism in American society. |
| Personal Reflections on Racism and Identity | racist, racism, white, black, think, people, I'm, you're, don't | Interpersonal and emotional exchanges on racial identity and accusations of racism. |
| Watchmen (HBO): White Supremacy and Policing | judd, angela, white night, keene, police, racist | Analysis of racialized violence and white supremacy themes in HBO's *Watchmen* narrative. |
| Historical Memory of the Tulsa Massacre and Racial Violence | tulsa, massacre, history, oklahoma, black, white, people, rene, jensen | Reflection on the Tulsa Race Massacre and how racial violence is remembered or forgotten in U.S. history. |

*Critical Discourse Analysis*

We employed Critical Discourse Analysis (CDA) to examine how language reflects and reproduces social power relations and ideological structures. In CDA, language is never neutral—it is analyzed for its ideological foundations and entanglement with broader sociocultural and political contexts. As Fairclough (2018) explains, CDA uncovers often opaque causal links between (a) discursive practices, events, and texts, and (b) the wider social and cultural structures that shape them. He conceptualizes discourse as a three-dimensional phenomenon: discourse events (actual texts), discursive practices (production, distribution, and consumption), and social practices (societal structures that inform and are informed by discourse).

    Using this framework, we analyzed our filtered dataset in NVivo. Each text was treated as a discourse event situated within discursive and social practices, allowing us to examine how

power, ideology, and identity are constructed and contested through language (Blommaert & Bulcaen, 2000). Through this analysis, we aimed to reveal how broader struggles over meaning and authority surface in language use.

In our study, Reddit threads about race served as discourse events. We analyzed their discursive practices—users' storytelling, argumentation, and positioning—and their social practices—how these interactions relate to broader narratives of race, identity, and media consumption. CDA enabled us to trace patterns in how users adopted roles such as advocates, adversaries, and adaptives, and the discursive strategies used to affirm, challenge, or complicate dominant narratives. By contextualizing these patterns within racialized media representation and networked publics, we show how Reddit becomes a space where popular culture and racial discourse intersect, enabling both the reinforcement and transformation of ideology.

**Findings**

This study examined how Reddit users discussed race in *Lovecraft Country* and *Watchmen*, identifying three fluid social roles: advocate, adversary, and adaptive. These roles were interactive and not mutually exclusive—individuals often shifted between them. For instance, someone might begin as an adversary, become adaptive after engaging with new perspectives, and eventually adopt advocacy. Roles also overlapped, such as when users supported one show but opposed the other. Advocates and adversaries often served as opinion leaders. Discursive sections further illustrate these dynamic role interactions.

*Reel Talk: Redditors Debate if Racism is Exaggerated or Real*

Drawing on Bandura's (1999) mechanisms of moral disengagement, online discussions about *Lovecraft Country* and *Watchmen* revealed how individuals displaced responsibility and minimized harm by downplaying racism's severity. This section explores how moral disengagement, advocacy, and adaptation manifested within Reddit debates about the shows' depictions of racial oppression. The Reddit discussions around the *Lovecraft Country* episode, *Sundown*, in which the show's three Black protagonists, Leti, George, and Tic, desperately attempt to escape a sundown town before sunset exemplify the interaction of social roles.

**Adversaries: Questioning the Realism of Racism Depictions.** Some Redditors, whom we term "adversaries," argued that the shows exaggerated racial violence for dramatic effect. Adversaries are individuals that engage in debates to oppose or challenge the shows' depictions or media messages as well as Redditors that showed support for the shows. One user commented that the depictions created a "hyperbolic version of the 1950s," suggesting it was unrealistic for Black Americans to experience such constant threats. These arguments exemplified strategies of minimizing historical trauma by reframing racism as rare rather than systemic. Bandura (1999) claims one tactic of moral evasion is to ignore, minimize, distort, or even dispute that one's activities cause any harm such as ignoring or diminishing the effects of racism. Disengagement may also include dehumanizing and blaming the victims for bringing harm to themselves or being deserving of the harm they have endured (Bandura, 1999, p. 3) as shown here, "It is hard to care about the main character or the oppression he faces because of what he has done. He deserves what he gets. I have no sympathy for him."

The above quote exemplifies how blaming victims of systemic injustice came in the form of deservedness based on the character's individual actions. By minimizing the frequency or realism in which covert racism persisted during Jim Crow and redirecting blame of systemic racism as an earned punishment to poor decisions and immoral actions reflect what Bonilla-Silva (2021) refers to as colorblindness. In which racism is justified on the basis of "character." In this sense, Tic, the character being identified here, deserves oppression based on character deficits which reduces feelings of sympathy or responsibility in the complicity of white supremacy.

**Advocates: Countering Disengagement Through Education.** In response, advocates challenged these narratives by sharing resources—historical documents, news links, and personal stories—that contextualized racial violence. Informants mobilized both fact-based and experience-based discourses, providing educational materials about sundown towns, lynchings, and systemic discrimination (Jakob, 2022; Carlson, 2016; Moe & Larsson, 2013). Rather than simply offering information, advocates framed their shared resources to guide interpretation. For instance, users posted links to historical overviews of sundown towns and personal testimonies connecting family histories to the events depicted. One user shared, "My grandfather had an experience like the one in that sundown town episode," while another described current-day racism tied to segregated spaces. These standpoint perspectives helped legitimize the narratives, using vicarious experiences to foster sympathy and engagement (Rajabi, 2021). One advocate exclaimed, "You just cannot relate because it wasn't your ancestors being burned, bombed, and murdered in the streets. You couldn't FEEL the narrative like we did."

**Adaptive Redditors: Learning and Shifting Perspectives.** A distinct group, adaptive Redditors, shifted from skepticism to acknowledgment after engaging with advocates' resources. Some initially doubted the scale of violence, stating disbelief that such lynchings or racial terror were widespread. However, after reading historical links or learning from others' experiences, they amended their views, noting, for example, "I wasn't aware that this kind of violence happened outside the South." Several adaptive Redditors expressed surprise upon learning that sundown towns and racial violence were not limited to the Southern United States. Resource-sharing about events like the Great Migration and massacres in Northern states (e.g., MOVE, when Philadelphia police fired 10,000 rounds of ammunition and bombed the residence of Black activists in 1985 leading to 11 deaths, including five children (Hall, 2024)) played a role in correcting misconceptions. As one user reflected, "I thought segregation was just a Southern thing—learning about sundown towns is eye-opening."

Adaptive Redditors sometimes crossed into the role of advocates themselves. Many appreciated the subreddit's educational function, noting, "I came here for entertainment but ended up learning so much history," or "Thanks to the resources posted, I now understand these historical references better." The evolution from adversary to adaptive participant highlights how Reddit's discussion format allowed users to reconsider their assumptions. Drawing on Habermas, Jakob (2022) explains that persuasion involves recognizing truth, moral rightness, and sincerity in discourse—processes clearly evident in how adaptive Redditors engaged with advocates' interventions (Warnke, 1995).

*#NotAllWhitePeople: Displacement and Diffusion of Responsibility*

*Lovecraft Country* and *Watchmen* use white characters as a symbol for and representation of white supremacy. In *Lovecraft Country,* the white characters are in juxtaposition of literal monsters (sometimes the white characters even turn into monsters), metaphorically highlighting the monstrosity of anti-Black racism (Lewis, 2021). Magic is used to symbolize privilege and power. The show ends with the Black protagonists, Leti, George, and Tic, removing the white villains' ability to use magic and generalize this effect so that no white person can use magic to oppress others in the future. *Lovecraft Country* has come under scrutiny from media academics and critics claiming that it may reinforce the very racial narratives it's meant to critique: through exploiting Black pain for profit it positions itself as "trauma porn" and that the supernatural elements obscures the very real terror of anti-Black racism (Phillips, 2020; Lewis, 2021).

**Adversaries: Racism against White People?** Some adversaries argued that *Lovecraft Country* unfairly portrayed all white characters as villains, suggesting the show exaggerated racial realities, "It felt like every white person the protagonists encountered were either hostile, terrified, or trying to harm them—almost cartoonishly so." Adversaries criticized *Lovecraft Country* creators stating, "the show is reinforcing the idea that race is an absolute. If the protagonists are celebrated for having a rigid definition and simplistic view of race, then the creators are promoting that view." Additionally, adversaries claimed that the broad stroke of removing all white magic denies the history of oppressed white ethnic groups such as Jewish, Irish, Greek, and Italian people and obscures the historical use of magic as rebellion against oppressors, especially by white women. They claimed that this is not an attempt at depicting equality, rather it is revenge.

**Advocates: Confronting Systemic Racism Through Discomfort.** Advocates countered that omitting "good" white characters was intentional, forcing white audiences to confront systemic racism without the comfort of identifying with heroic figures. As one advocate explained, "Having a 'good white character' would let white viewers excuse themselves, thinking they'd be one of the few who stood against racism. The discomfort is the point." This narrative choice resists Hollywood's typical portrayal of white saviors, where white protagonists dominate films about racism, sidelining Black activism and framing racial inequality as an individual moral flaw rather than a systemic issue (Larson, 2006; Kellner, 2011).

Advocates praised eliminating the white magic narrative encouraging others to imagine the possibilities of Black power. One advocate said, "imagine if Malcolm X had magical powers, and Civil Rights leaders, the world would be a much different place. Imagine that Black people could use magic to influence laws and politicians, think of all the possibilities!" Other advocates criticized adversaries' positions, saying adversaries are using "race is a social construct" as a scapegoat and urged them to just accept that the show may not be for them. Advocates reinforced this by highlighting that magic is a metaphor for power wielded by oppressors and that many Black Americans have had to watch countless movies and shows in which violence against Black people is glorified. These advocates requested that the adversaries stop using red herrings of reverse racism and posited that adversaries just wanted a "white-friendly," happy ending.

**In-between Adversaries and Advocates.** Like *Lovecraft Country*, *Watchmen* also centers Black heroes and white villains. Interestingly, some adversaries criticized *Lovecraft Country* as heavy-handed on racism while praising *Watchmen* for what they perceived as a more "nuanced" portrayal of racism. This highlights the fluidity of Reddit users' social roles—users could simultaneously critique one show and support another. One user who was adverse toward *Lovecraft Country*, embraced the racial discourse in *Watchmen*, even promoting *Watchmen* as a better educational resource for learning about race. The user was much less resistant to the racial commentary provided in *Watchmen*, but did express resistance to the more overt depictions of racism in *Lovecraft Country*. When narratives are perceived as overly didactic or exaggerated, audience resistance increases, leading to reactance—scrutiny and counter-arguing, a well-documented phenomenon in resistance to narrative persuasion (Moyer-Gusé & Nabi, 2010). This suggests that *Watchmen's* "nuanced" portrayals were more palatable content for some. However, based on whether someone found *Lovecraft Country* or *Watchmen* palatable was utilized by others as a gauge to identify covert racism.

**Adaptives: Accepting New Positions.** One adaptive Redditor who began as an adversary expressed their discomfort with the white characters depiction as immoral in *Lovecraft Country*. They stated, "I'm not from the US, but I thought American racism was mostly old laws like Black people must go in a different door to a business or had to sit at the back of the bus. But that was in the past." After receiving multiple comments and links to resources from advocates, the Redditor shifted from adversary to adaptive thanking commentors for the resources, stating, "I learned many upsetting and disturbing things." Additionally, one advocate shifted to an adaptive position agreeing with an adversary who argued for more nuanced portrayals of race, especially narrative plots in *Lovecraft Country* that exclude all white people from magic. This adaptive Redditor stated, "I agree I hope future seasons will show more nuanced depictions of race." This highlights how Redditors may adopt or accept advocate or adversary positions.

**Using *Lovecraft Country* and *Watchmen* as "Racism Detectors."** Discussions questioned whether disliking the shows indicated latent racism. Some users reported being accused of racism for criticizing narrative choices, "I didn't love how over-the-top the cop scenes were, and suddenly I'm labeled a racist." Others admitted that the shows stirred discomfort among people close to them, "A friend quit watching after one episode, saying he wasn't interested in a show about Black people standing up to racists. The show really surfaced hidden biases."

In a study looking at how college students interpreted anonymous posts online, Eschmann (2020), stated that individuals who had a more critical lens and deeper understanding of race relations, deciphered different meanings of the posts; whereas some individuals did not detect masked racism or were ignorant or agnostic toward issues of race, standpoint experiences allowed a keener detection of microaggressions and masked racism. Advocates argued that dismissing the shows' portrayal of racism often stemmed from racial ignorance, "It's not written to make white viewers comfortable—it's designed to challenge their role in this history."

Some Redditors rejected racial commentary altogether, framing their criticisms as a desire to avoid "politics" in entertainment, "I'm just tired of every show I watch lately being about racism. Sometimes you just want to escape." This resistance reflects a broader reluctance to engage with media that confronts racial realities head-on. We also found a resistance to racial narratives through canonical fidelity as discussed below.

*Canonical Fidelity as Covert Resistance*

Discussions around narrative authenticity and fidelity to canon emerged as a recurring flashpoint in Reddit threads, often functioning as a discursive proxy for discomfort with racialized storytelling. While some users explicitly voiced critiques of the shows' racial themes, others invoked claims of inauthenticity—questioning whether *Lovecraft Country* and *Watchmen* "really" stayed true to their source material or genre conventions. This subsection explores how such appeals to canonical fidelity masked ideological resistance. Drawing on Proctor's (2019) notion of fans as "textual conservationists," we analyze how adversaries framed racial storytelling as a deviation from narrative legitimacy, using genre expectations as a rhetorical shield. At the same time, advocates interrogated these critiques, revealing patterns of selective nostalgia and racial exclusion embedded in canon-policing discourse.

**Adversaries: Protecting the Storyline.** Proctor (2019) describes fans as "textual conservationists" who resist narrative changes under the guise of protecting canon. Critiques of *Lovecraft Country* and *Watchmen* sometimes masked discomfort with racial themes behind complaints about authenticity, "This show isn't really Lovecraftian horror—it feels like a different story altogether." In some cases, canonical complaints overlapped with anti-woke sentiment, suggesting that arguments about story authenticity served as a coded rejection of racialized storytelling, "The original *Watchmen* criticized all politics. The HBO version feels like it's just pushing woke identity politics." Thus, canonical fidelity complaints may serve as vehicles for covert racism, revealing discomfort with centering Black history and trauma.

**Advocates: Using Reddit's Affordances to Contextualize Criticisms.** Advocates challenged these canonical critiques, noting that some users appeared less interested in textual fidelity than in disrupting conversations about Black-centered narratives, "Instead of letting Black fans enjoy the show, they come here to loudly declare it's 'objectively bad.'" Advocates also leveraged Reddit's affordances, examining adversaries' posting histories to expose political affiliations, "Whenever someone complains about the 'exaggerated racism,' I check their account—and often find a long history of posts in conservative subreddits." This practice highlights how Redditors contextualized users' criticisms within broader ideological patterns.

*Les Ripoux: Rotten Cops and the Persistence of Racism*

The portrayal of policing in *Lovecraft Country* and *Watchmen* prompted contentious debates among Reddit users, illuminating stark divides in perceptions of law enforcement and systemic violence. Police were depicted not as neutral enforcers of justice but as emblematic of racial terror, prompting both backlash and affirmation. This section examines how Redditors engaged with the trope of the "rotten cop"—or les ripoux—and how this narrative challenged long-standing media representations of police as heroic protectors. While adversaries mobilized the "few bad apples" defense and reframed critiques of policing as attacks on law and order, advocates drew connections between the shows' portrayals and real-world events such as the murder of George Floyd and the BLM protests. These tensions surfaced deeper ideological rifts around accountability, institutional trust, and the role of media in interrogating systemic racism.

**Advocates: Fiction Reflects Reality.** Advocates emphasized that racism depicted in *Lovecraft Country* and *Watchmen* mirrors persistent realities, citing the 2020 Black Lives Matter protests as evidence that systemic racism endures. Police portrayal, in particular, sparked significant debate. As Bernardo (2022) notes, prior to BLM, television often depicted police as benevolent protectors, even legitimizing extralegal violence. However, since BLM, media depictions have become more critical, reflecting shifting public sentiment. Both shows challenge the traditional "hero cop" narrative, presenting police as morally ambiguous or openly corrupt. Discussions among Redditors reflected a real-world binary view of police—as protectors of democracy or as authoritarian enforcers of racial hierarchy. One user commented, "Thinking police departments will actually hold bad cops accountable is a fantasy." Advocates often framed unchecked policing as a form of state-sanctioned vigilantism, echoing *Watchmen*'s original critique of power with the phrase, "who watches the watchmen?" Some Redditors wrestled with their discomfort, noting that while authoritarianism is dangerous, racist terrorism is even worse—forcing uneasy support for some *Watchmen* characters.

**Adversaries: The "Few Bad Apples" Defense.** Adversaries invoked the "few bad apples" narrative, arguing that condemning all police was unfair. Some also reproduced white racial framing, blaming violence on Black culture rather than systemic oppression, and claimed media coverage of BLM protests incited division. In response, advocates drew sharp connections between the shows' fictional portrayals and real-world injustices: "Abuse of power by cops isn't rare—it's woven into the system." Debates extended to historical events referenced in the shows, such as the Trumbull Park Riots (1953-1966) and Red Summer. Advocates shared articles and resources documenting systemic police brutality during these events, while adversaries argued that shows like *Lovecraft Country* unfairly generalized all police as racist.

## *Connecting Past and Present Racism*

Reddit discussions revealed how users connected historical atrocities to modern incidents. One user reflected, "Nice white folks might not have lynched anyone themselves, but ignoring injustice allowed it to thrive—same as today, when people excuse violence against Black Americans." Others noted the continued relevance of racial violence, "Today's headlines—church attacks, gated communities barring Black youth—mirror scenes from the 1950s." These comments underscore how viewers linked the shows' fictionalized racism to persistent real-world patterns. This reinforces the argument that the past is not as distant as some would like to believe as emphasized by one user, "Racism today isn't gone; it's just less visible. Black people are still being killed by police and attacked in churches."

**Conclusion**

This study examined how Reddit audiences engaged with the racial narratives of *Lovecraft Country* and *Watchmen*, showing how discussions of race, history, and power are shaped through fluid social roles and platform affordances. Redditors did not hold fixed ideological positions; instead, they often shifted between advocate, adversary, and adaptive roles, reflecting ongoing negotiations of racial meaning. Debates around the realism of racism portrayals, fidelity to source material, and divergent views on police narratives revealed broader dynamics of moral disengagement (Bandura, 1999, 2012), colorblind racial ideology (Bonilla-Silva, 2021), and resistance to narrative persuasion. Adversaries expressed discomfort through claims of exaggeration, reverse racism, or appeals for apolitical entertainment. In contrast, advocates responded with standpoint experiences, historical evidence (Jakob, 2022; Carlson, 2016; Moe & Larsson, 2013), and emotional appeals grounded in vicarious experience (Kaplan, 2005; Rajabi, 2021).

Reddit's pseudonymity enabled both covert racism and the emergence of issue publics that challenged dominant narratives (Massanari, 2015; Warner, 2002). Many discussions linked fictional content to real-world injustices, such as suppressed Black histories, police violence, and systemic white supremacy (Bernardo, 2022; Sturken, 2007; Delgado & Stefancic, 2023). Rather than inducing guilt, advocates emphasized awareness of historical and ongoing injustice. Notably, some users moved from adversary to adaptive roles, suggesting that even adversarial online spaces can support racial learning and reflection. These findings contribute to ongoing discussions about the role of popular culture, digital publics, and participatory media in the contemporary struggle over racial meaning and historical memory.

This paper makes three key theoretical contributions. First, it extends narrative persuasion to pseudonymous digital publics, demonstrating that persuasive storytelling is not confined to traditional broadcast or interpersonal settings. On platforms like Reddit, narrative persuasion operates through dialogic and participatory processes, where users engage with media texts and with each other, co-constructing meaning in comment threads and discussions. This finding highlights how storytelling functions dynamically in decentralized, user-driven environments, where persuasion is as much about interaction as it is about content.

Second, it advances multistep flow theory in networked environments by highlighting how influence circulates via emergent opinion leaders—advocates and adversaries—who emerge with recognizable pseudonyms. These figures reinterpret and guide discourse in the absence of formal authority, revealing the algorithmically mediated and horizontal nature of online influence.

Third, it introduces a typology of fluid social roles in online racial discourse: advocate, adversary, and adaptive. This heuristic highlights how individuals shift positions through engagement with others' lived experiences, standpoint storytelling, and collective reasoning. In doing so, it reveals the transformative potential of digital discourse, where ideological change can emerge through tension, contestation, and solidarity.

This study highlights how Reddit audiences engaging with *Lovecraft Country* and *Watchmen* participate in dynamic, role-shifting discourse around race, memory, and power—embodying advocate, adversary, and adaptive positions rather than fixed ideological camps. These fluid interactions underscore how pseudonymous, countercultural platforms like Reddit afford both authentic expressions and harmful evasions, as Renninger (2014) suggests, by enabling users to "be themselves" while masking accountability. The discursive landscape is further shaped by broader sociotechnical dynamics of identity, fandom, and digital propaganda, as Daniels (2012) argues, where epistemic authority is contested and often racialized. Rather than treating online misrecognitions as mere "post-truth" confusion, our findings—alongside recent work in political epistemology (Walters et al., 2024) and critical race discourse—point to more deliberate struggles over how race is known, felt, and debated in popular culture. These racial reckonings are not peripheral—they are central to the social functions of participatory platforms, where interpretive communities co-construct meaning, police belonging, and model or resist transformation. Our findings show that platforms like Reddit serve not just as sites of entertainment commentary, but as arenas of ideological production, where cultural memory and racial politics are constantly in flux.

*Limitations and Future Research*

Despite its contributions, this study has several limitations. First, the analysis is confined to two television shows within specific subreddits. As such, the findings may not generalize to other genres, platforms, or communities with different moderation norms or user demographics. Second, while pseudonymous identities were analyzed through discourse, we did not track longitudinal participation patterns or employ digital ethnography to explore identity persistence across threads or subreddits. Third, our study focused on English-language Reddit discussions, potentially excluding racial narratives articulated in other cultural or linguistic contexts.

Future research could explore how users navigate race-centered media in more diverse digital ecologies, such as TikTok or Instagram, or how platform moderation policies impact the trajectory of racial discourse. Additionally, comparative studies across national contexts could reveal how historical memory and racial reckoning manifest differently in global digital publics.